\documentclass[lettersize,journal]{IEEEtran}
\usepackage{amsmath,amsfonts}
\usepackage{algorithmic}
\usepackage{array}
\usepackage{textcomp}
\usepackage{stfloats}
\usepackage{url}
\usepackage{verbatim}
\usepackage{graphicx}
\usepackage{cite}
\usepackage{booktabs}
\usepackage{multirow}
\usepackage{caption}
\usepackage{subcaption}
\def\BibTeX{{\rm B\kern-.05em{\sc i\kern-.025em b}\kern-.08em
    T\kern-.1667em\lower.7ex\hbox{E}\kern-.125emX}}
\usepackage{balance}
\begin{document}
\title{Automatic Disfluency Detection  \\
from Untranscribed Speech}
\author{Amrit Romana, 
Kazuhito Koishida, Emily Mower Provost
\thanks{A. Romana and E. Mower Provost are with University of Michigan, Ann Arbor, Michigan, USA.}
\thanks{K. Koishida is with Microsoft Corporation, Redmond, Washington, USA}
}

\maketitle

\begin{abstract}
Speech disfluencies, such as filled pauses or repetitions, are disruptions in the typical flow of speech. Stuttering is a speech disorder characterized by a high rate of disfluencies, but all individuals speak with some disfluencies and the rates of disfluencies may by increased by factors such as cognitive load. Clinically, automatic disfluency detection may help in treatment planning for individuals who stutter. Outside of the clinic, automatic disfluency detection may serve as a pre-processing step to improve natural language understanding in downstream applications. With this wide range of applications in mind, we investigate language, acoustic, and multimodal methods for frame-level automatic disfluency detection and categorization. Each of these methods relies on audio as an input. First, we evaluate several automatic speech recognition (ASR) systems in terms of their ability to transcribe disfluencies, measured using disfluency error rates. We then use these ASR transcripts as input to a language-based disfluency detection model. We find that disfluency detection performance is largely limited by the quality of transcripts and alignments. We find that an acoustic-based approach that does not require transcription as an intermediate step outperforms the ASR language approach. Finally, we present multimodal architectures which we find improve disfluency detection performance over the unimodal approaches. Ultimately, this work introduces novel approaches for automatic frame-level disfluency and categorization. In the long term, this will help researchers incorporate automatic disfluency detection into a range of applications. 
\end{abstract}


\section{Introduction}
\label{sec:intro}

A speech disfluency is a disruption in the typical flow of speech. Common disfluencies include filled pauses, repetitions, and revisions. Detecting, categorizing, and localizing disfluencies has several clinical and non-clinical applications. For example, previous work has found that disfluencies can provide insight into the language planning process and cognitive load~\cite{corley2008hesitation, lindstrom2008effect, van2014classification}, suggesting that automatic disfluency detection and categorization may help us understand how disfluencies relate to cognition \cite{romana2021automatically}. As another application, clinicians measure disfluency types and durations in planning treatment for individuals who stutter~\cite{yaruss1997clinical, yaruss2006overall, riley2009ssi}, and disfluency categorization and localization could make this treatment planning more scalable~\cite{riad2020identification}. Finally, outside of the clinic, previous work has found that disfluency detection and localization can be used as a pre-processing step to improve natural language understanding in downstream applications (e.g. with voice assistants or speech dictation systems)~\cite{arjun2020automatic, mitra2021analysis, lea2021sep, lea2023user}. With these applications in mind, we present new techniques for automatic disfluency detection, categorization, and localization.

The majority of work in disfluency detection has taken a language-based approach. Most recently, large language models achieved state-of-the-art results in this field \cite{rocholl2021disfluency, romana2022enabling}. In these works, the authors illustrate how BERT (Bidirectional Encoder Representations from Transformers, \cite{devlin2019bert}) can be fine-tuned to detect, for each token in a transcript, whether or not it is disfluent. However, these methods are not scalable due to their reliance on manually transcribed text. Automatic speech recognition (ASR) can efficiently generate text, but these methods currently lag in accuracy for atypical speech.

\begin{table}[t]
    \centering
    \caption{Examples of each disfluency class. In this paper, we aim to detect and categorize the audio frames associated with these disfluency classes, with examples underlined. Note: a token can belong to multiple classes, for example, the partial word shown here is labeled as a partial word and a repetition.}
    \resizebox{0.43\textwidth}{!}{%
    \begin{tabular}{ll}
    \toprule
    Filled pause (FP) &  And \underline{um} I think one thing...\\
    Partial word (PW) & \underline{H-} how do you feel about that? \\ 
    Repetition (RP) & well \underline{with my} with my grandmother...\\
    Revision (RV) & And uh \underline{we were} I was fortunate...\\
    Restart (RS) & \underline{If you} how long do you want to stay?\\
    \bottomrule
    \end{tabular}
    }
    \label{tab:disf_example}
\end{table} 

\begin{figure}[t]
    \centering
    \includegraphics[scale=0.65]{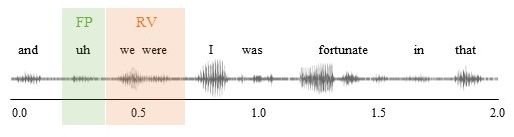}
    \caption{Disfluency detection, categorization, and localization task: We focus not only on detecting which segments have disfluencies or which tokens are disfluent, but locating the audio frames associated with each disfluency.}
    \label{fig:taskillustration}
\end{figure}

Using language introduced in \cite{lea2023user}, we define two opposing ASR goals for disfluent speech: \textit{intended} versus \textit{verbatim} speech transcription. Lea et al. focused on \textit{intended speech} transcription, which removes disfluencies and only transcribes the speakers' intentions~\cite{lea2023user}. As an example, if a user articulated ``and uh we were I was fortunate...,'' an intended speech transcript would drop disfluencies and only include ``and I was fortunate...'' Intended speech transcription is useful for applications when disfluencies impede understanding of the speaker's intent, but it does not allow for disfluency detection, categorization, or localization because the disfluencies are not transcribed. Other applications, for example clinical measurements of disfluencies, require \textit{verbatim speech} transcription~\cite{romana2021automatically, romana2023multimodal}. Using the same example text, a verbatim speech transcript would contain everything that was articulated, including disfluencies: ``and uh we were I was fortunate...'' Verbatim speech transcripts allow for an analysis of disfluencies, and could be combined with disfluency detection methods to derive intended speech transcripts. In this work, we provide the first comparison of different ASR systems for verbatim speech transcription and a novel approach for downstream disfluency detection, categorization, and localization.  

While language-based approaches have been most commonly explored for this task, modern acoustic approaches have also shown potential. For example, Wav2vec 2.0 (W2V2) is a self-supervised speech representation model commonly used for speech recognition \cite{baevski2020wav2vec} and a small body of work has begun to analyze the use of W2V2 in disfluency detection\cite{bayerl2022detecting, romana2023multimodal}. In this work, we compare the use of W2V2 with more recent acoustic-based speech representation models for disfluency detection. 

We evaluate different approaches for frame-level automatic disfluency detection and categorization. We start by comparing how different ASR systems perform in verbatim speech transcription. We then use these transcripts as input to a language-based disfluency detection model, and we evaluate how the models' performance is limited by the quality of ASR transcripts. Next, we investigate disfluency detection from acoustic signals where we omit the intermediate transcription step. We find that the acoustic-only approach offers an improvement over the ASR-language approach for most disfluency classes. Finally, we present new multimodal architectures that offer an additional improvement in performance over the ASR-language and acoustic-only approaches. 

We focus on the categorization of the disfluency types listed in Table \ref{tab:disf_example}, where these classes are largely based on those defined in early disfluency work~\cite{johnson1961measurements} and still used by clinicians and researchers today. We complete our analysis at the frame-level, as is illustrated in Figure \ref{fig:taskillustration} and has been done in \cite{romana2023multimodal, riad2020identification}. Frame-level detection, or localization, has two primary benefits over word- or segment-level detection for the applications described: 1) frame-level predictions can easily be mapped to disfluency durations which have clinical value \cite{yaruss1997clinical, riley2009ssi}, and 2) outside of the clinic, frame-level predictions can indicate which frames should be masked to improve downstream tasks \cite{mendelev2021improved}. In addition, frame-level detection allows us to compare performance across a range of input types, including manual transcripts, ASR transcripts, and raw audio signals, which can all be mapped to the frame-level. 

The novelty of this work is:
\begin{itemize}
    \item An analysis of how different ASR systems, including open-source versus proprietary options, as well as fine-tuned versus off-the-shelf options, compare in generating verbatim speech transcripts. 
    \item An evaluation of how different ASR systems perform when their transcripts are used for downstream disfluency detection. 
    \item An evaluation of how different acoustic representations can be leveraged for detecting disfluencies without transcription as an intermediate step. 
    \item A presentation of multimodal architectures, and a detailed comparison of how a multimodal approach improves performance over a language or acoustic one. 
    \item The release of our model weights with demos, making it easy to evaluate how the language, acoustic, and multimodal methods perform on other datasets\footnote{\url{github.com/amritkromana/disfluency_detection_from_audio}}.
\end{itemize}

\noindent Ultimately, this work is an important step toward automatic disfluency detection and in the long-term a wide range of speech applications would benefit from the incorporation of these methods. 

\section{Related Work}

In this section, we describe previous work in disfluency detection and categorization. We start by describing language-based efforts that process manually transcribed text. With scalability in mind, we then discuss work that has explored the impact of ASR on disfluencies. To avoid potential ASR bottlenecks, we then describe acoustic-based efforts that do not require transcription. And finally, we describe related work that has combined acoustic and language features. 

\subsection{Manually transcribed text and disfluency detection}

The majority of work in disfluency detection and categorization has taken a language-based approach focused on tagging words in manually transcribed text. Initially, these works focused on extracting handcrafted features associated with each word, including part-of-speech tags and distance pattern matching metrics (such as if a word or phrase equals the following word or phrase). First, researchers used these features with conditional random fields (CRFs) to classify disfluent words~\cite{georgila2009using, ostendorf2013sequential, zayats2014multi}. With the introduction of recurrent neural networks (RNNs), Zayats et al.~proposed the use of a bi-directional long short-term memory (BLSTM) network that processed handcrafted features~\cite{zayats2016disfluency}. 

Next, researchers moved towards leveraging pre-trained word embeddings. Pre-trained word embeddings have been shown to encode token characteristics better than handcrafted features for a variety of tasks~\cite{wei2019twitter, chen2018end}. Bach et al.~proposed a BLSTM that took pre-trained character and word embeddings as input for disfluency detection~\cite{bach2019noisy}. %Their model was trained with additional noise to increase robustness. 
Lou et al.~proposed an auto-correlational neural network (ACNN) that took pre-trained word embeddings as input~\cite{lou2018disfluency}. Their method describes disfluencies (specifically repetitions or revisions) and their corrections as ``rough copies'' of each other. They then use auto-correlation to identify these ``rough copies.'' The introduction of self-attention has led to further improvements. Self-attention relates different positions within a sequence and also could be used to find these ``rough copies.'' Leveraging self-attention, Wang et al.~introduced a state-of-the-art language-based disfluency detection approach based on transformers~\cite{wang2018semi, wang2020multi}.  

Most recently, researchers have shifted from training randomly initialized neural networks to fine-tuning pre-trained language models, such as BERT (Bidirectional Encoder Representations from Transformers)~\cite{devlin2019bert}. These models benefit from learning general characteristics from large quantities of data and multiple pre-training tasks and have shown success when fine-tuned for a wide range of NLP tasks~\cite{tsai2019small, gao2019target, adoma2020comparative}. 
Rocholl et al.~illustrated that a fine-tuned BERT model outperformed previously used CRFs and LSTMs in detecting disfluencies~\cite{rocholl2021disfluency}. Romana et al.~expanded on this analysis and found that a fine-tuned BERT model could be used for both disfluency detection and categorization, and that these models generalized well to pathological speech~\cite{romana2022enabling}. 

These language-based methods have largely been evaluated with manually transcribed text. In our earlier work, we illustrated how the performance of these methods drops significantly when ASR transcripts are used as inputs~\cite{romana2023multimodal}. One limitation of that work is that we only evaluated one ASR system, so in this work we compare disfluency detection performance using manual transcripts and a range of ASR systems. 

\subsection{ASR text and disfluency detection}

Researchers have started to evaluate how ASR performs with disfluent speech and how this may impact downstream tasks. Overall, they found that current ASR systems, which perform well with typical, fluent speech, underperform for disfluent speech. However, the next direction that ASR systems should take depends on the downstream task. We refer to these two directions as verbatim speech transcription versus intended speech transcription, and each direction is useful for certain applications. On one hand, clinicians require verbatim transcripts where disfluencies are transcribed so that they can be evaluated~\cite{yaruss1997clinical, riley2009ssi}. On the other hand, dictation services benefit from transcribing only the intended, or nondisfluent, speech~\cite{mitra2021analysis}. 

Romana et al. investigated ASR for verbatim transcription~\cite{romana2021automatically}. This work demonstrated that disfluencies may be a predictor of MoCA score (a screening tool for cognitive impairment, \cite{nasreddine2005montreal}) in patient's with Parkinson's Disease. They then explored if ASR transcripts generated by Deepspeech \cite{hannun2014deep} could be used as input in this MoCA prediction task. In moving from manually transcribed text to ASR text, they found that ASR systems would delete disfluencies and these errors propagated to the downstream MoCA score prediction model, thus limiting the use of disfluency related features. 

Lea et al. focused on intended speech transcription. They explored how people who stutter, where stuttering is characterized by an increase in disfluencies, interact with voice assistants and dictation services \cite{lea2023user}. These services rely on ASR and Lea et al. find that, for these services, individuals preferred to only see their intended speech transcribed. They evaluated Apple Speech framework \cite{applespeechframework} and find that it does not sufficiently delete disfluencies and this creates a barrier for using these technologies.

These works illustrate how current ASR systems have not been trained specifically for verbatim or intended speech transcription. With disfluency detection in mind, we need to first evaluate how these methods perform for verbatim speech transcription. In our paper, we compare how different ASR systems perform in verbatim transcription and we evaluate how these transcripts compare when used as input for disfluency detection. We leave an analysis of intended speech transcription to future work.  

\subsection{Acoustic-based approaches for disfluency detection}

Previous work has found that disfluencies exhibit distinct acoustic cues~\cite{savova2003prosodic, moniz2012prosodic}. Acoustic approaches for disfluency detection benefit because they do not need to rely transcripts as input and as a result they would not be impacted by ASR errors. %However, only a small body of work has explored acoustic approaches for this task. 

Rachid et al. introduced a set of handcrafted acoustic features, including word duration, pause times, and statistics on fundamental frequency and energy~\cite{riad2020identification}. Using speech from people who stutter, they illustrated that these features could be used in a support vector machine model to detect and categorize a broad set of disfluencies. This analysis was evaluated at the frame-level. However, the authors found that the accuracy of these handcrafted acoustic features significantly lagged behind previously introduced language features. 

Lea et al.~expanded on the use of these acoustic features, but rather than extracting features at the word-level (ex. the mean energy or fundamental frequency within a word), they model the full time-series of acoustic features (ex. mel-filter banks and fundamental frequency over time in a segment)~\cite{lea2021sep}. With this modification, their approach did not rely on transcripts at all. They introduced a convolutional LSTM (ConvLSTM), an architecture beginning with a CNN for feature extraction and then using an LSTM to process the features and complete the final classification at the segment-level.

As with the language approaches, large pre-trained models that generate acoustic representations, such as those used for ASR, have shown promise for disfluency detection and categorization. Bayerl et al.~found that a fine-tuned W2V2 model outperformed the previously introduced ConvLSTM for detecting and categorizing disfluencies at the segment-level~\cite{bayerl2022detecting}. In our earlier work, we finetuned W2V2 for acoustic-based frame-level disfluency detection and categorization ~\cite{romana2023multimodal}. One limitation of that work is that we only evaluated W2V2, so in this work we compare W2V2 to other models that we hypothesize may generalize better to this non-ASR task. In this paper, we also provide a comparison between the improved ASR-language and acoustic approaches. 

\subsection{Multimodal approaches for disfluency detection}

Language and acoustic approaches have been useful in different ways for disfluency detection and categorization but only a small body of work has explored multimodal approaches for this task. A few studies have augmented language-based approaches with prosody features corresponding to each word~\cite{shriberg1997prosody, shriberg1999phonetic, ferguson2015disfluency, zayats2016disfluency}. For example, Shriberg et al.~introduced a decision tree model that combined language and prosody for detecting and categorizing filled pauses, repetitions, and revisions at the word-level~\cite{shriberg1997prosody}. They found that the multimodal approach outperformed the two unimodal ones, and within the prosody features, word duration was of highest importance. As another example, Zayats and Ostendorf incorporated prosody features in their language-based classifier and found that this improved the model's performance over a language-only approach~\cite{zayats2019giving}. Their architecture concatenates the outputs of two LSTMs: one for processing text features (including word embeddings and part of speech tags) and another for processing prosody features (including pauses, energy, and fundamental frequency within a word). Their proposed model then classifies each word as disfluent or fluent. 

These multimodal methods do not leverage the pre-trained language or acoustic models that have demonstrated state-of-the-art performance for unimodal approaches. Further, these methods rely on word-level features and they have only been explored with manually transcribed text. It is unclear how well these methods would perform when this text is not available and needs to be generated with ASR. In our earlier work, we introduced a multimodal model combining language representations from a fine-tuned BERT model and acoustic representations from a fine-tuned W2V2 model, and we found that this multimodal approach improved over the unimodal ones~\cite{romana2023multimodal}. We evaluated this approach with language representations from ASR text, and found that the multimodal approach was especially beneficial here as it could compensate for ASR errors. In this work, we expand upon this multimodal analysis by leveraging improved language and acoustic representations, evaluating different fusion architectures, and providing a detailed comparison of how each modality performs.   

\section{Dataset}
\label{sec:dataset}

The Switchboard corpus is a widely used dataset of telephone conversations~\cite{godfrey1992switchboard}. We analyze the audio provided by the data curators. These recordings were sampled at 8 kHz. We upsample the audio to 16 kHz as is required for inputs to the various models investigated in this paper and is commonly done \cite{yang2022torchaudio, chan2021speechstew, salesky2021multilingual}. We use the Mississippi State University (MSU) transcripts, which includes transcript corrections and more accurate word alignments, which are important for frame-level detection~\cite{deshmukh1998resegmentation}. We use the segment boundaries and outside-error-correction disfluency labels provided by Zayats et al~\cite{zayats2019disfluencies}. These labels correspond to the tokens in the MSU transcripts, and indicate if the token is an error, a correction, or outside (as in, not part of an error or a correction). We drop a 3,556 segments (1.9\% of total segments) where there were disagreements between the labels, text, and text timings files. We add 50 ms of silence to the beginning and end of the audio for the included segments, and we convert the outside-error-correction labels to more detailed disfluency classes listed in Table \ref{tab:disf_example}. These disfluencies classes are important for clinical metrics and allow for more fine-grained understanding of how speakers struggle with specific disfluencies. We apply a rule-based approach, from our prior work \cite{romana2022enabling, romana2023multimodal}, to complete the mapping from the provided disfluency labels to these classes. We label occurrences of ``um'' and ``uh'' as filled pauses. We label words ending with ``-'' (such as ``th-'') as partial words. Given the provided error and correction annotations, we label an error as a repetition if it matches its correction, a revision if it does not match its correction, and a restart if there is no corresponding correction. For this paper, we make one change from the labeling approach used in our earlier work: we allow a word to have multiple labels. Namely, a token can be described as a partial word as well as a repetition or revision, allowing for the differentiation between a partial word repetition or a partial word revision. After defining these labels, we normalize the transcripts by making the text lower case and removing punctuation from this list: . ? ! , -.  We release the data preprocessing code to encourage others to contribute to frame-level disfluency detection and categorization.

\begin{figure}
    \centering
    \includegraphics[scale=0.5]{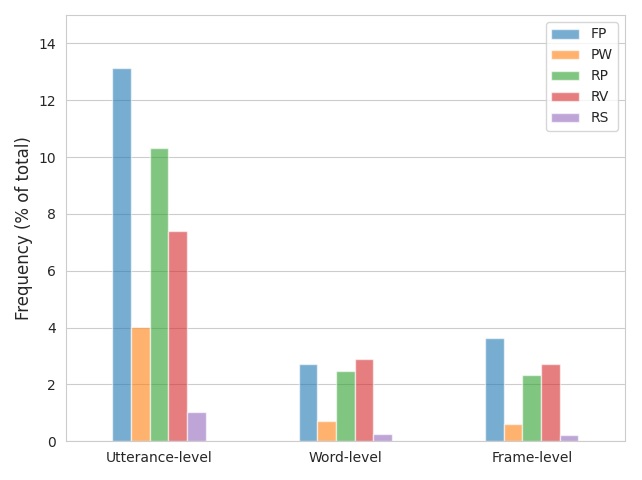}
    \caption{Proportion of utterances, words, and frames that contain each type of disfluency: filled pauses (FP), partial words (PW), repetitions (RP), revisions (RV), and restarts (RS).}
    \label{fig:datafreq}
\end{figure}

Figure \ref{fig:datafreq} displays the frequency of each class at the utterance-, word-, and frame-level. Analysis at the word- or frame-level is subject to higher class imbalances than utterance-level analysis, and partial words and restarts are particularly rare. 

We use the train, development, test splits established in previous work~\cite{charniak2001edit}. All file names begin with ``sw'' followed by a 4 digit conversation ID. File names beginning with ``sw2'' or ``sw3'' are included in the train set (167,777 segments). File names beginning with ``sw45,'' ``sw46,'' ``sw47,'' ``sw48,'' or ``sw49'' are included in the development set (9,722 segments). File names that begin with ``sw40'' or ``sw41'' are included in the test set (7,529 segments).
%TRAIN=170921
%#VAL=9971
%TEST=7702 
%TOTAL=188584 BEFORE DROP
%3556 were dropped 
%NEW TOTAL IS 185028

\begin{figure*}
    \centering
    \includegraphics[width=\textwidth]{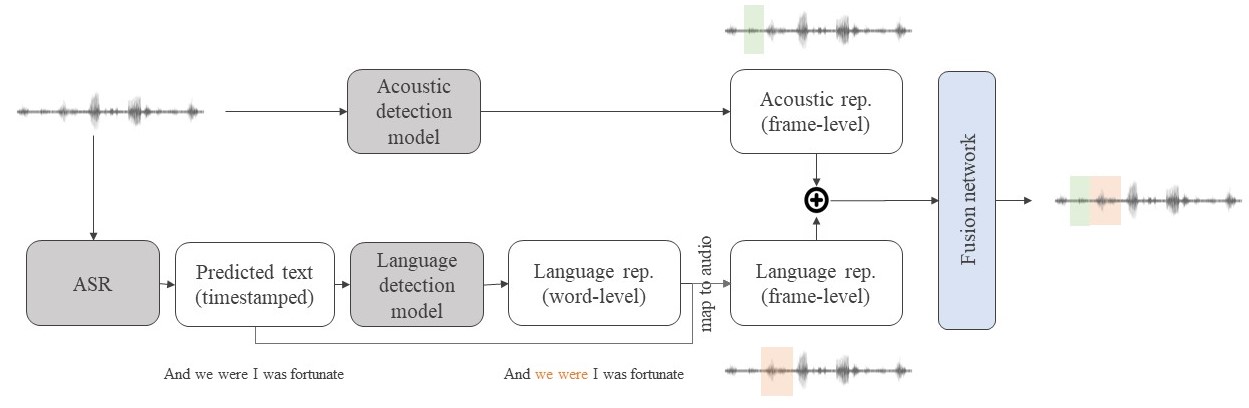}
    \caption{Multimodal fusion. Audio are input to a fine-tuned acoustic model (WavLM) for acoustic representations of disfluencies. Audio are also input to ASR (Whisper-FT) which produces timestamped transcripts. The text are input to a fine-tuned language model (BERT) for language representations of disfluencies. We use the timestamps to upsample these word-level representations to the frame-level, concatenate the acoustic and language representations, and then train a fusion network.}
    \label{fig:diagram}
\end{figure*}

\section{Methodology}   
\label{sec:method}

We compare several approaches for disfluency detection. We start with a best-case scenario: manually transcribed text is available during training and testing. However, the  impracticality of manually transcribing a large number of utterances makes this scenario difficult to scale. Next, we consider a more practical scenario: manually transcribed text is available during training of the model, but not during testing. For this case, we evaluate how performance of the language-based disfluency detection model changes when ASR transcripts are used during testing. To avoid the bottleneck of ASR transcription, we also consider if audio-alone could be used for disfluency detection. Finally, we present a multimodal approach that combines the unimodal representations. In this section, we first describe the ASR systems used to derive transcripts and the metrics we use to compare them. Then we describe the disfluency detection models analyzed. We implement this work using the PyTorch and HuggingFace libraries~\cite{paszke2019pytorch, wolf-etal-2020-transformers}

\subsection{ASR transcripts}
\label{sec:method_asr}

ASR systems are typically not trained to transcribe disfluencies and thus the transcripts may not be reliable for language-based disfluency detection. In this section, we describe a range of modern systems we use to extract transcripts. Then, we quantify how different ASR systems may perform for this task. We compare their performance in terms of word error rate (WER) and frame error rate (FER). Additionally, we break down WER and FER to their disfluent and nondisfluent components and analyze the performance of each ASR system across these metrics.

\textbf{Self-supervised models: W2V2, HuBERT, and WavLM}

W2V2, HuBERT, and WavLM are open-source systems that have risen in popularity due to the fact that they have been pre-trained on large quantities of unlabeled data and only need to be fine-tuned with small quantities of data. These systems first encode speech audio via a series of convolutional layers and then pass these encodings through a transformer network to build contextualized representations. W2V2 jointly learns a discrete vocabulary of latent speech representations, which is used to train the model with a contrastive loss. After pretraining on unlabeled speech, W2V2 can be fine-tuned with a smaller quantity of labeled data and Connectionist Temporal Classification (CTC) loss~\cite{baevski2020wav2vec, graves2006connectionist}. HuBERT builds on W2V2 but is pre-trained using a cross-entropy loss and the targets are built during a separate clustering process. As a result, HuBERT learns better representations and achieves lower WER than W2V2~\cite{hsu2021hubert}. WavLM builds on HuBERT but is pre-trained with an additional denoising task and leverages an adaptive positional encoding. As a result, WavLM achives lower WER than HuBERT. WavLM also outperforms BERT on non-ASR tasks, such as speech separation and speaker diarization~\cite{chen2022wavlm}. 

These models show potential for disfluency detection in two ways: we can use the ASR-generated transcripts in a language-based disfluency detection model, or we could use the contextualized speech representations as input to an acoustic-based disfluency detection model. In this section we focus on the former language-based approach, and we discuss the acoustic-based approach in Section \ref{sec:acoustic}. 

We evaluate the use of these ASR systems, off-the-shelf (OTS) and fine-tuned (FT), in generating transcripts for a language-based disfluency detection model. For the OTS models, we download the large-sized W2V2, HuBERT, and WavLM checkpoints that have been pre-trained with the 960 hour Librispeech dataset and fine-tuned with a 100 hour labeled subset \cite{panayotov2015librispeech}. We decode these models using a CTC decoder and a 5-gram language model built on the same subset of Librispeech \cite{heafield2011kenlm}. For the FT model experiments, we build a 5-gram language model on the Switchboard training data. We download the same large-sized checkpoints, freeze the convolutional feature encoder, and fine-tune the transformer only (315 million parameters) with the Switchboard training data, as was recommended in \cite{baevski2020wav2vec}. During training, we exclude audio shorter than 0.5 seconds (due to insufficient audio) and longer than 10 seconds (due to resource constraints). We train the model using an Adam optimizer, CTC loss, and the following hyperparameters: an effective batch size of 128, a maximum learning rate of 1e-4 with 1000 warmup steps and linear decay, evaluation every 500 steps, and a patience of 5 where the best model was chosen based on lowest loss on the development set. This was achieved with 11,500 training steps (9.0 epochs) for W2V2, 10,000 training steps (7.8 epochs) for HuBERT, and 12,500 training steps (9.7 epochs) for WavLM. We use the fine-tuned models, the CTC decoder, and the language model to derive timestamps for the final transcripts. 

\textbf{Weakly-supervised model: Whisper}

More recently, Radford et al. introduced Whisper as a state-of-the-art open-source ASR system \cite{radford2022robust}. Whisper takes input audio, converts it into a log-Mel spectrogram, and then passes these features into an encoder-decoder transformer. It has been pre-trained on a wide variety of tasks: language identification, phrase-level timestamps generation, multilingual speech transcription, and to-English speech translation. The pre-training used 680,000 hours of diverse data collected from the web, so Whisper does not typically require fine-tuning for domain specific transcription. In our work, we use the Whisper-small checkpoint which contains 240 million trainable parameters so it is most comparable to the size of the self-supervised models described in the previous section. 

We evaluate the use of Whisper off-the-shelf (OTS) and fine-tuned (FT), in generating transcripts for a language-based disfluency detection model. While Whisper-OTS performs well for transcribing nondisfluent words, its performance is more inline with intended speech transcription (as in, it does not transcribe disfluencies) and this is reflected in our results. Whisper-OTS does include a decoding flag to detect disfluencies, but disfluencies are only transcribed as ``*'' using this method. We evaluated how well the predicted timestamps for ``*'' compared with the Switchboard frame-level disfluencies, and we found a recall score of 0.21. 

We hypothesize that if Whisper is fine-tuned with verbatim disfluent speech, its transcripts can be used in a language-based disfluency detection model with a higher average recall score. We fine-tune the Whisper checkpoint using the Switchboard training data. We limit the training data to audio up to 10 seconds for this step. We train the model using an Adam optimizer, CTC loss, and the following hyperparameters: an effective batch size of 128, a maximum learning rate of 1e-5 with 1000 warmup steps and linear decay, evaluation every 500 steps, and a patience of 5 where the best model was chosen based on lowest loss. This was achieved with 2,500 training steps (2.0 epochs). We derive timestamps for the final transcriptions using dynamic time warping (DTW) \cite{JSSv031i07}. Specifically, we drop the language modeling head and apply DTW to cross-attention weights using the method provided by Linagora Lab \cite{lintoai2023whispertimestamped}. 

\textbf{Proprietary system: Azure}

Finally, we compare the use of these open-source models to a proprietary option: Microsoft Azure speech-to-text service (Azure). We use the Azure REST API to extract transcripts with timestamps. Our earlier work \cite{romana2023multimodal} focused on the use of these transcripts as they could accurately be extracted in an off-the-shelf manner. One limitation of this method is that the API service requires a subscription and in deriving these transcripts we do not gain insight into the inner workings of the model.

\textbf{Disfluency Error Rates}

\begin{table*}[t]
    \centering
    
    \caption{ASR Results for disfluency transcription and timestamping. To compare transcription accuracies, we present overall word error rate (WER), as well as disfluent WER (WER-D) and nondisfluent WER (WER-ND). To evaluate the timestamp accuracies, we present frame error rate (FER), as well as disfluent FER (FER-D) and nondisfluent FER (FER-ND). The best results for each metric are in bold. 
    }
    \begin{tabular}{c | ccc | ccc}
    \toprule
    ASR System & WER & WER-ND & WER-D & FER & FER-ND & FER-D\\
    \midrule
    W2V2-OTS & 0.40 & 0.17 & 0.66 & 0.45 & 0.42 & 0.83 \\
    W2V2-FT & 0.09 & 0.05 & 0.18 & 0.37 & 0.34 & 0.68 \\
    HuBERT-OTS & 0.47 & 0.20 & 0.66 & 0.49 & 0.46 & 0.84 \\
    HuBERT-FT & 0.10 & 0.05 & 0.20 & 0.36 & 0.33 & 0.70 \\
    WavLM-OTS & 0.33 & 0.13 & 0.57 & 0.41 & 0.37 & 0.80 \\
    WavLM-FT & \textbf{0.08} & 0.04 & \textbf{0.16} & \textbf{0.35} & \textbf{0.32} & 0.66 \\
    \midrule
    Whisper-OTS & 0.19 & 0.05 & 0.71 & 0.54 & 0.52 & 0.74 \\
    Whisper-FT & \textbf{0.08} & \textbf{0.03} & \textbf{0.16} & 0.42 & 0.43 & \textbf{0.36} \\
    \midrule
    Azure-OTS & 0.11 & 0.04 & 0.25 & \textbf{0.35} & 0.33 & 0.47 \\
    \bottomrule
    \end{tabular} 
\label{tab:asr_results}
\end{table*}

We standardize the ASR transcripts using the same process described for the manual transcripts in Section \ref{sec:dataset}. We also address common spelling changes, such as ``umm'' versus ``um,'' ``uhm'' versus ``um,'' and ``ok'' versus ``okay.'' We define new error metrics that quantify how well each of these methods performs in transcribing not only words overall, but transcribing and locating nondisfluent and disfluent words.  

Because the ground truth and predicted transcripts may contain different numbers of words, we rely on the Levenshtein distance algorithm to determine where words have been inserted, substituted, or deleted. We consider the number of words ($nwords$), number of insertions ($I$), number of deletions ($D$), and number of substitutions ($S$).  Then, we calculate the word error rate (WER) as: 

\begin{equation}
WER = \frac{I+D+S}{nwords}
\label{eq:wer}
\end{equation}

Using a binary disfluent tag associated with each word, we count the number of disfluent words ($nwords_{d}$) and number of nondisfluentwords ($nwords_{n})$ where $nwords=nwords_{d}+nwords_{n}$. Additionally, we count the number of disfluent and nondisfluent deletions ($D_{d}$ and $D_{n}$, respectively), number of disfluent and nondisfluent substitutuions ($S_{d}$ and $S_{n}$, respectively), and number of nondisfluent insertions ($I_{n}$). We observe that insertions tend to be nondisfluent (e.g., ASR systems will not artificially insert a repetition or a filled pause), and so we do not consider the insertion of disfluent words. With these counts defined, we calculate a nondisfluent WER (WER-ND) and a disfluent WER (WER-D) as: 

\begin{equation}
WER{\text -}ND = \frac{I_{n} + D_{n} + S_{n}}{nwords_{n}}
\end{equation}

\begin{equation}
WER{\text -}D = \frac{D_{d} + S_{d}}{nwords_{d}}    
\end{equation}

We need to evaluate how well these methods identify the timestamps corresponding with each word which allows us to compare the frame-level accuracy of manual versus ASR transcripts. We copy each word from the ground truth and predicted transcripts to each frame it spans or is predicted to span, and then we calculate a frame error rate (FER). Because the number of frames in the audio does not depend on the transcribed text, (i.e., the length of the audio does not change) we do not need a distance metric such as that we used for the WER. Instead, for a set of ASR predictions, we can calculate, for each frame, whether it contains the correct word or a transcription error. We count the total number of frames ($nframes$) and the total number of frames with errors ($nframes_{e}$). We then use these counts to calculate a frame error rate (FER):  

\begin{equation}
FER = \frac{nframes_{e}}{nframes}    
\end{equation}

Finally, we label frames as disfluent or nondisfluent depending on if there is a disfluent word in that frame or not according to the ground truth labels and we count the number of frames with and without disfluent words ($nframes_{d}$ and $nframes_{n}$, respectively). We also count the number of frames with transcription errors and with or without disfluent words ($nframes_{e,d}$ and $nframes_{e,n}$, respectively). With these counts, we calculate a nondisfluent FER (FER-ND) and disfluent FER (FER-D):

\begin{equation}
FER{\text -}ND = \frac{nframes_{e,n}}{nframes_{n}} 
\end{equation}

\begin{equation}
FER{\text -}D = \frac{nframes_{e,d}}{nframes_{d}}
\end{equation}

Table \ref{tab:asr_results} shows these metrics on the Switchboard test set across the different systems analyzed. For some applications (such as with dictation services or voice assistants), we may prefer to only transcribe the intended, or nondisfluent, speech. Whisper-OTS yields a low WER (19\%) and the biggest difference between the WER-ND and WER-D (5\% versus 71\%). It also benefits from being open-source and OTS. As a result, it is likely the most promising candidate for intended speech transcription. With other applications (such as in clinical measurements of disfluencies), disfluencies need to be transcribed verbatim and aligned with audio. FT models tend to have better performance than their OTS counterparts for those applications. We find WavLM-FT, Whisper-FT, and Azure-OTS perform the best. WavLM-FT and Whisper-FT achieve the lowest WER (8\%) and WER-D (16\%). Azure-OTS performs slightly worse in terms of WER (11\%) and WER-D (25\%). At the frame-level, WavLM-FT and Azure-OTS achieve the lowest FER (35\%). However, Whisper-FT achieves the lowest FER-D (36\%) while FER-D is higher for WavLM-FT and Azure-OTS (66\% and 47\%, respectively). In the rest of this section, we will explore how WavLM-FT, Whisper-FT, and Azure-OTS perform as inputs to a language-based disfluency detection model. We will also introduce acoustic- and multimodal approaches for this task.  

\subsection{Frame-level Disfluency Detection and Categorization}

In this section, we describe the different approaches we compare for frame-level disfluency detection and categorization. We start with a language-based model and evaluate it using both manually-created and ASR-generated transcripts as input during testing. We also evaluate acoustic-based models that are equally as scalable as the ASR transcript approach but avoid the transcription bottleneck. Finally, we present a multimodal approach that combines the ASR language and acoustic representations at the frame-level and offers an improvement in performance over the unimodal approaches. 

In implementing each of these models, we need to account for the different time scales at which they operate. While the language model outputs a prediction for each token, the acoustic models outputs predictions for each frame, where frames are 25 ms long with a 20 ms shift. Thus, we need to generate corresponding frame-level labels. We copy the word-level labels to each frame that the word spans given the timestamps associated with the word. We present all final metrics (F1 and recall score) at the frame-level to have a consistent comparison across the different methods. More specifically, this frame-level evaluation allows us to 1) compare performance of the language model with manual versus ASR transcripts which may contain different tokens and timestamps and 2) compare the performance of the language and acoustic methods which operate at different time scales. 

\textbf{A language-based approach} \label{sec:bert}

We fine-tune the BERT language model using the manually transcribed text as input and the word-level disfluency labels as targets. We start with the pre-trained BERT base model (uncased), append a linear output layer, and fine-tune the entire model (the pre-trained encoder as well as the output layer). We train with a batch size of 64 and use the Adam optimizer to minimize binary cross entropy loss. We use the Switchboard development set to optimize the learning rate (1e-4, 5e-5, 1e-5) and number of training steps (up to 15 epochs). We evaluate performance on the development set every 100 training steps, and we choose the best model based on unweighted average recall (UAR) and a patience of 50 steps. We repeat this training using 3 random seeds, and we report average and standard deviation of metrics. 

We evaluate these models on the manually transcribed text for a best case analysis. However, relying on manually transcribed text limits the scalability of this approach. ASR can be efficiently used to generate text but, in our earlier work ~\cite{romana2023multimodal}, we illustrated how disfluency detection performance deteriorated when ASR transcripts were used as input to the fine-tuned language-based disfluency classifier. In this work, we expand upon this analysis by comparing the use of WavLM-FT, Whisper-FT, and Azure-OTS, which we found in Section \ref{sec:method_asr} performed well for disfluency transcription. Note that we only use these ASR transcripts during test-time, and not during fine-tuning, because the language model processes word-level disfluency labels and these word-level labels correspond to words in the ground truth transcripts.  

Because the manually and ASR transcribed text may contain different words, word-level disfluency detection metrics are difficult to compare. We present the final metrics at the frame-level. We use the word start and end timings from the manual or ASR transcripts to upsample the word-level disfluency predictions to the frame-level. We calculate the final metrics by comparing these frame-level predictions to frame-level labels, and we present these results in Section \ref{sec:results}.

\textbf{An acoustic-based approach} \label{sec:acoustic}

We compare the use of W2V2, HuBERT, and WavLM as acoustic-based disfluency detection models. Rather than using the acoustic models for  ASR and using the transcripts as inputs to a language-based model, we use the acoustic models' speech representations. In our earlier work, we found that W2V2 could be fine-tuned for acoustic-based disfluency detection \cite{romana2023multimodal}. HuBERT and WavLM output representations at the same-frequency (25 ms long frames with a 20 ms shift), but these newer models have been pre-trained in ways that they may be more generalizeable to non-ASR tasks \cite{hsu2021hubert, chen2022wavlm}. 

We append an output layer to the base models for W2V2, HuBERT, and WavLM. We freeze the convolutional layers and fine-tune the transformer and output layers as was recommended in \cite{baevski2020wav2vec}. Each of these models outputs frame-level predictions, so we train and evaluate this model with the frame-level disfluency labels. We train with a batch size of 64 and use the Adam optimizer to minimize binary cross entropy loss. We use the Switchboard development set to optimize the learning rate (1e-4, 5e-5, 1e-5) and number of training steps (up to 15 epochs). We evaluate performance on the development set every 100 training steps, and we choose the best model based on unweighted average recall (UAR) and a patience of 50 steps. We repeat this training using 3 random seeds, and we report average and standard deviation of metrics for each model. 

\textbf{A multimodal approach} 

We hypothesize that a multimodal fusion model will improve detection over the unimodal approaches. The multimodal model will be able to leverage the fact that language representations better capture the semantics of the text compared to the acoustic representations. At the same time, the multimodal model will be able to leverage the acoustic representations to compensate for transcription or alignment errors that were introduced by the ASR system and propagated through to the language representations. 

Our multimodal approaches takes language and acoustic representations, concatenated at the frame-level, as input. First, we drop the output layer of the fine-tuned BERT model, and use the resulting model to derive 768-dim language representations from ASR transcripts. 
Next, we upsample these language representations to the frame-level, using the method we used to upsample ground truth and predicted labels. We set the language representation to a vector of zeros for frames that do not contain a word. 
Then, we drop the output layer from the fine-tuned acoustic model, and use the resulting model to derive 768-dim acoustic representations from audio. The upsampling step standardized the scale for the language and acoustic representations, and we concatenate the two to generate a 1536-dim frame-level representation. We use these frame-level representations as input to our multimodal fusion networks.

We present three multimodal fusion networks: a single-layer perceptron, a bidirectional long short-term memory network (BLSTM), and a transformer network. A simple single-layer perceptron may suffice because the input representations for each frame have already been contextualized with the other frames. However, a BLSTM, which processes inputs sequentially forward and backward, may better model the frame-level representations. More so, transformers benefit from parallelizing the sequential computations. A transformer network may also improve performance with its attention mechanism. 

For the single-layer perceptron, we initialize the model weights with the weights from the output layers of the acoustic and language models, and we fine-tune it with the Switchboard training set. We use an Adam optimizer and the Switchboard development set to optimize the learning rate (1e-4, 5e-5, 1e-5) and number of training steps (up to 10 epochs). For the BLSTM, we randomly initialize a network with 1 layer and a hidden size of 512 as was done in our earlier work \cite{romana2023multimodal}. This network has 8.4 million parameters to train. For the transformer network, we randomly initialize a transformer encoder with 1 layer, 8 attention heads, and a feedforward layer with a dimension of 1024. This results in a network with 12.6 million parameters to train. We train these models with the Switchboard training set. We use an Adam optimizer with the hyperparameters and learning rate scheduler specified in \cite{vaswani2017attention} and implemented by OpenNMT \cite{opennmt}. Specifically, for the optimizer, $\beta_1=0.9$, $\beta_2=0.98$, $\epsilon=10^{-9}$, and the learning rate ($lr$) varies according to a schedule: 
$lr=d^{-0.5}*min(stepnum^{-0.5}, stepnum*warmupsteps^{-1.5})$ 
where $d$ is the size of the model. We use Switchboard development set to optimize the number of warmup steps (500, 1000, 2000) and number of training steps (up to 10 epochs). For all models, we evaluate performance on the development set every 100 training steps, and we choose the best model based on unweighted average recall (UAR) and a patience of 50 steps. We repeat this training using 3 random seeds, and we report average and standard deviation of metrics. 

%three types of classifier: a single-layer perceptron, a bi-directional long short term memory network (BLSTM), and a transformer model. The first model, a single-layer perceptron is the simplest model, but, given that the input representations for each frame have already been contextualized with the other frames, we hypothesize that a simple model may suffice and have the benefit of high interpretability. To speed up the training, and reduce variation across runs, we initialize the dense layer with the pre-trained layer weights from the unimodel classifiers. The second model, a BLSTM, was evaluated to be effective in combining results, 

%To evaluate this, we train another multimodal fusion model where we generate language representations for our ASR transcripts from the fine-tuned BERT model described in Section \ref{sec:bert}. We then repeat the upsampling, concatenation with W2V2 representations, and BLSTM training to predict disfluency labels for each frame. This process is illustrated in Figure \ref{fig:diagram}.

\begin{table*}[t]
    \caption{Disfluency detection and categorization performance using a fine-tuned BERT model with different \textbf{language} inputs: ground truth transcripts versus transcripts from different ASR systems: WavLM-FT, Whisper-FT, and Azure-OTS, where FT=fine-tuned and OTS=off-the-shelf. The target disfluency classes include filled pauses (FP), repetitions (RP), revisions (RV), restarts (RS), and partial words (PW). The best ASR results for each metric are in bold. 
    }
    \begin{subtable}[h]{\textwidth}
        \centering
        \caption{F1 score}
        \begin{tabular}{c | cc | ccccc | c}
        \toprule
        & \multicolumn{2}{c |}{Disfluency Macros} & \multicolumn{5}{c | }{Disfluency Classes} & Non-\\
        Transcript type & Unweighted & Weighted & FP & RP & RV & RS & PW & Disfluent \\
        \midrule
        Manual & 0.70 (0.00) & 0.85 (0.00) & 1.00 (0.00) & 0.85 (0.01) & 0.70 (0.01) & 0.13 (0.01) & 0.83 (0.01) & 0.99 (0.00)\\
        \midrule
        WavLM - FT & 0.32 (0.01) & 0.45 (0.01) & 0.52 (0.00) & 0.49 (0.01) & 0.40 (0.01) & 0.06 (0.01) & 0.12 (0.01) & \textbf{0.97 (0.00)}\\
        Whisper - FT & 0.35 (0.01) & 0.47 (0.01) & 0.56 (0.00) & 0.48 (0.02) & 0.45 (0.01) & 0.07 (0.01) & \textbf{0.17 (0.01)} & 0.95 (0.00) \\
        Azure - OTS & \textbf{0.38 (0.00)} & \textbf{0.51 (0.00)} & \textbf{0.58 (0.01)} & \textbf{0.56 (0.01)} & \textbf{0.50 (0.00)} & \textbf{0.10 (0.01)} & \textbf{0.17 (0.01)} & 0.96 (0.00) \\
        \bottomrule
        \end{tabular} 
        \label{tab:asr_f1score}
    \end{subtable}
    \newline
    \vspace*{10 pt}
    \newline
    \begin{subtable}[h]{\textwidth}
        \centering
        \caption{Recall score}
        \begin{tabular}{c | cc | ccccc | c}
        \toprule
        & \multicolumn{2}{c |}{Disfluency Macros} & \multicolumn{5}{c | }{Disfluency Classes} & Non-\\
        Transcript type & Unweighted & Weighted & FP & RP & RV & RS & PW & Disfluent \\
        \midrule
        Manual & 0.71 (0.01) & 0.86 (0.00) & 1.00 (0.00) & 0.88 (0.00) & 0.68 (0.00) & 0.09 (0.00) & 0.88 (0.02) & 0.99 (0.00)\\
        \midrule
        WavLM - FT & 0.26 (0.00) & 0.36 (0.00) & 0.40 (0.00) & 0.40 (0.01) & 0.33 (0.00) & 0.04 (0.00) & 0.13 (0.01) & \textbf{0.98 (0.00)}\\
        Whisper - FT & \textbf{0.44 (0.00)} & \textbf{0.60 (0.00)} & \textbf{0.76 (0.00)} & \textbf{0.62 (0.01)} & 0.48 (0.00) & 0.05 (0.01) & \textbf{0.28 (0.01)} & 0.93 (0.00) \\
        Azure - OTS & 0.40 (0.00) & 0.56 (0.00) & 0.67 (0.01) & \textbf{0.62 (0.01)} & \textbf{0.50 (0.00)} & \textbf{0.06 (0.01)} & 0.13 (0.00) & 0.96 (0.00) \\
        \bottomrule
        \end{tabular} 
        \label{tab:asr_recallscore}
    \end{subtable}
\end{table*}

\begin{table*}[t]
    \caption{Disfluency detection and categorization performance using different \textbf{acoustic} models: W2V2, HuBERT, and WavLM, which have all been fine-tuned for the task. The target disfluency classes include filled pauses (FP), repetitions (RP), revisions (RV), restarts (RS), and partial words (PW). The best results for each metric are in bold. 
    }
    \begin{subtable}[h]{\textwidth}
        \centering
        \caption{F1 score}
        \begin{tabular}{c | cc | ccccc | c}
        \toprule
        & \multicolumn{2}{c |}{Disfluency Macros} & \multicolumn{5}{c | }{Disfluency Classes} & Non-\\
        Acoustic model & Unweighted & Weighted & FP & RP & RV & RS & PW & Disfluent \\
        \midrule
        W2V2 & 0.39 (0.01) & 0.56 (0.01) & 0.86 (0.00) & 0.52 (0.02) & 0.30 (0.04) & 0.00 (0.00) & 0.29 (0.04) & \textbf{0.98 (0.00})\\
        HuBERT & 0.42 (0.00) & 0.60 (0.01) & \textbf{0.88 (0.01)} & 0.55 (0.02) & 0.35 (0.01) & 0.00 (0.00) & 0.31 (0.03) & \textbf{0.98 (0.00)}\\
        WavLM & \textbf{0.45 (0.01)} & \textbf{0.64 (0.01)} & 0.87 (0.01) & \textbf{0.64 (0.01)} & \textbf{0.44 (0.04)} & 0.00 (0.00) & \textbf{0.32 (0.01)} & \textbf{0.98 (0.00)}\\
        \bottomrule
        \end{tabular} 
    \label{tab:acous_f1score}
    \end{subtable}
    \newline
    \vspace*{10 pt}
    \newline
    \begin{subtable}[h]{\textwidth}
        \centering
        \caption{Recall score}
        \begin{tabular}{c | cc | ccccc | c}
        \toprule
        & \multicolumn{2}{c |}{Disfluency Macros} & \multicolumn{5}{c | }{Disfluency Classes} & Non-\\
        Acoustic model & Unweighted & Weighted & FP & RP & RV & RS & PW & Disfluent \\
        \midrule
        W2V2 & 0.42 (0.00) & 0.59 (0.00) & 0.85 (0.01) & 0.57 (0.01) & 0.36 (0.02) & 0.00 (0.00) & \textbf{0.33 (0.02)} & \textbf{0.98 (0.00)}\\
        HuBERT & 0.42 (0.00) & 0.60 (0.01) & \textbf{0.88 (0.01)} & 0.55 (0.02) & 0.35 (0.01) & 0.00 (0.00) & 0.31 (0.03) & \textbf{0.98 (0.00)}\\
        WavLM & \textbf{0.45 (0.01)} & \textbf{0.64 (0.01)} & 0.87 (0.01) & \textbf{0.64 (0.01)} & \textbf{0.44 (0.04)} & 0.00 (0.00) & 0.32 (0.01) & \textbf{0.98 (0.00)}\\
        \bottomrule
        \end{tabular} 
    \label{tab:acous_recallscore}
    \end{subtable}
\end{table*}

\section{Results}
\label{sec:results}

\subsection{Language-based approaches}

Tables \ref{tab:asr_f1score} and \ref{tab:asr_recallscore} list the disfluency detection and categorization results on the Switchboard test set using manually versus ASR transcribed text. 

Using manually transcribed text, the BERT classifier achieves high accuracy for most disfluency classes. Filled pauses can be detected with a recall score of 1.00. Repetitions and partial words are slighty more difficult, but these can still be detected with recall scores of 0.88 each. Revisions, which require increased context understanding, can be detected with a recall score of 0.68, and restarts which both require increased context and suffer due to limited samples, can only be detected with a recall score of 0.09. The F1 scores show a pattern similar to recall scores. 
 
In the more practical scenario where ASR transcripts are used as input, we see large performance drops. These drops can be attributed to both transcription errors as well as alignment errors. Whisper-FT performs the best in terms of recall score for most disfluency types. But when comparing the use of manual transcripts to Whisper-FT, unweighted and weighted recall averages drop from 0.71 to 0.26 and 0.86 to 0.36, respectively. The drop is largest for partial words, where recall score with manual transcripts is 0.88 but with Whisper-FT transcripts is 0.28. Azure-OTS leads to lower recall scores than Whisper-FT (unweighted recall of 0.40 versus 0.44) but higher F1 scores (unweighted F1 of 0.38 versus 0.35). We suspect this is because, as shown in Table \ref{tab:asr_results}, Whisper-FT has a lower FER-D than Azure-OTS (36\% versus 47\%) but it has a higher FER-ND than Azure-OTS (43\% versus 33\%). WavLM-FT performs much worse than Whisper-FT and Azure-OTS, and we suspect this is directly related to its relatively high FER-D (66\%).  

These results leave the question, would automatic disfluency detection benefit from skipping transcription and operating directly on audio?  

\subsection{Acoustic-based approaches} 

Tables \ref{tab:acous_f1score} and \ref{tab:acous_recallscore} list the disfluency detection and categorization results on the Switchboard test set using acoustic models. Comparing W2V2, HuBERT, and WavLM, we find that WavLM performs best in terms of F1 and recall score macros. The largest differences that we see are for repetitions and revisions. For repetitions, recall increases from 0.57 or 0.55 with W2V2 and HuBERT, respectively, to 0.64 with WavLM. For revisions, recall increases from 0.36 or 0.35 with W2V2 and HuBERT, respectively, to 0.44 with WavLM. We expect that this is a result of pre-training choices that were made for WavLM. Specifically, it was the only model trained on a non-speech recognition task and as a result it is better generalizing to the task at hand. 

When comparing the performance of this acoustic-only WavLM to the ASR systems introduced in the previous section, we find that the WavLM achieves better performance than the ASR systems for filled pauses and partial words, it performs comparably for repetitions, and it performs worse for revisions and restarts. We suspect this is due to the limited semantic understanding of this acoustic-based model. Next, we explore if combining the language and acoustic representations may improve model performance.

\begin{table*}[t]
    \caption{Disfluency detection and categorization performance using \textbf{fusion networks} that couple acoustic representations (from the WavLM acoustic model) and automatically-derived language representations (from the fine-tuned BERT model with Whisper-FT transcripts as input). The target disfluency classes include filled pauses (FP), repetitions (RP), revisions (RV), restarts (RS), and partial words (PW). The best results for each metric are in bold.
    }
    \begin{subtable}[h]{\textwidth}
        \centering
        \caption{F1 score}
        \begin{tabular}{c | cc | ccccc | c}
        \toprule
        & \multicolumn{2}{c |}{Disfluency Macros} & \multicolumn{5}{c | }{Disfluency Classes} & Non-\\
        Approach & Unweighted & Weighted & FP & RP & RV & RS & PW & Disfluent \\
        \midrule
        ASR-Language: Whisper-FT & 0.35 (0.01) & 0.47 (0.01) & 0.56 (0.00) & 0.48 (0.02) & 0.45 (0.01) & 0.07 (0.01) & 0.17 (0.01) & 0.95 (0.00) \\
        Acoustic: WavLM & 0.45 (0.01) & 0.64 (0.01) & \textbf{0.87 (0.01)} & 0.64 (0.01) & 0.44 (0.04) & 0.00 (0.00) & 0.32 (0.01) & 0.98 (0.00)\\
        \midrule
        Multimodal: Perceptron & 0.49 (0.01) & 0.64 (0.01) & 0.80 (0.03) & 0.65 (0.01) & 0.52 (0.01) & 0.08 (0.02) & 0.37 (0.00) & \textbf{0.98 (0.00)} \\ 
        Multimodal: BLSTM & \textbf{0.52 (0.01)} & \textbf{0.69 (0.01)} & 0.86 (0.00) & \textbf{0.69 (0.02)} & \textbf{0.55 (0.02)} & \textbf{0.09 (0.01)} & \textbf{0.40 (0.01)} & \textbf{0.98 (0.00)} \\ 
        Multimodal: Transformer & 0.50 (0.01) & 0.67 (0.01) & 0.86 (0.00) & 0.68 (0.01) & 0.51 (0.01) & 0.06 (0.03) & 0.39 (0.01) & \textbf{0.98 (0.00)} \\ 
        \bottomrule
        \end{tabular} 
        \label{tab:mm_f1score}
    \end{subtable}
    \newline
    \vspace*{10 pt}
    \newline
    \begin{subtable}[h]{\textwidth}
        \centering
        \caption{Recall score}
        \begin{tabular}{c | cc | ccccc | c}
        \toprule
        & \multicolumn{2}{c |}{Disfluency Macros} & \multicolumn{5}{c | }{Disfluency Classes} & Non-\\
       Approach & Unweighted & Weighted & FP & RP & RV & RS & PW & Disfluent \\
        \midrule
        ASR-Language: Whisper-FT & 0.44 (0.00) & 0.60 (0.00) & 0.76 (0.00) & 0.62 (0.01) & 0.48 (0.00) & 0.05 (0.01) & 0.28 (0.01) & 0.93 (0.00) \\
        Acoustic: WavLM & 0.45 (0.01) & 0.64 (0.01) & 0.87 (0.01) & 0.64 (0.01) & 0.44 (0.04) & 0.00 (0.00) & 0.32 (0.01) & 0.98 (0.00)\\
        \midrule
        Multimodal: Perceptron & 0.44 (0.00) & 0.61 (0.01) & 0.80 (0.03) & 0.61 (0.01) & 0.47 (0.01) & 0.05 (0.02) & 0.29 (0.01) & 0.98 (0.00) \\ 
        Multimodal: BLSTM & \textbf{0.48 (0.00)} & \textbf{0.66 (0.01)} & \textbf{0.88 (0.01)} & 0.66 (0.02) & \textbf{0.50 (0.04)} & \textbf{0.06 (0.00)} & 0.33 (0.04) & \textbf{0.99 (0.00)} \\ 
        Multimodal: Transformer & 0.47 (0.00) & 0.64 (0.00) & 0.86 (0.02) & \textbf{0.67 (0.01)} & 0.43 (0.01) & 0.03 (0.02) & \textbf{0.35 (0.04)} & \textbf{0.99 (0.00)} \\ 
        \bottomrule
        \end{tabular} 
    \label{tab:mm_recallscore}
    \end{subtable}
\end{table*}

\begin{figure*}[h]
    \centering
    \includegraphics[scale=0.3]{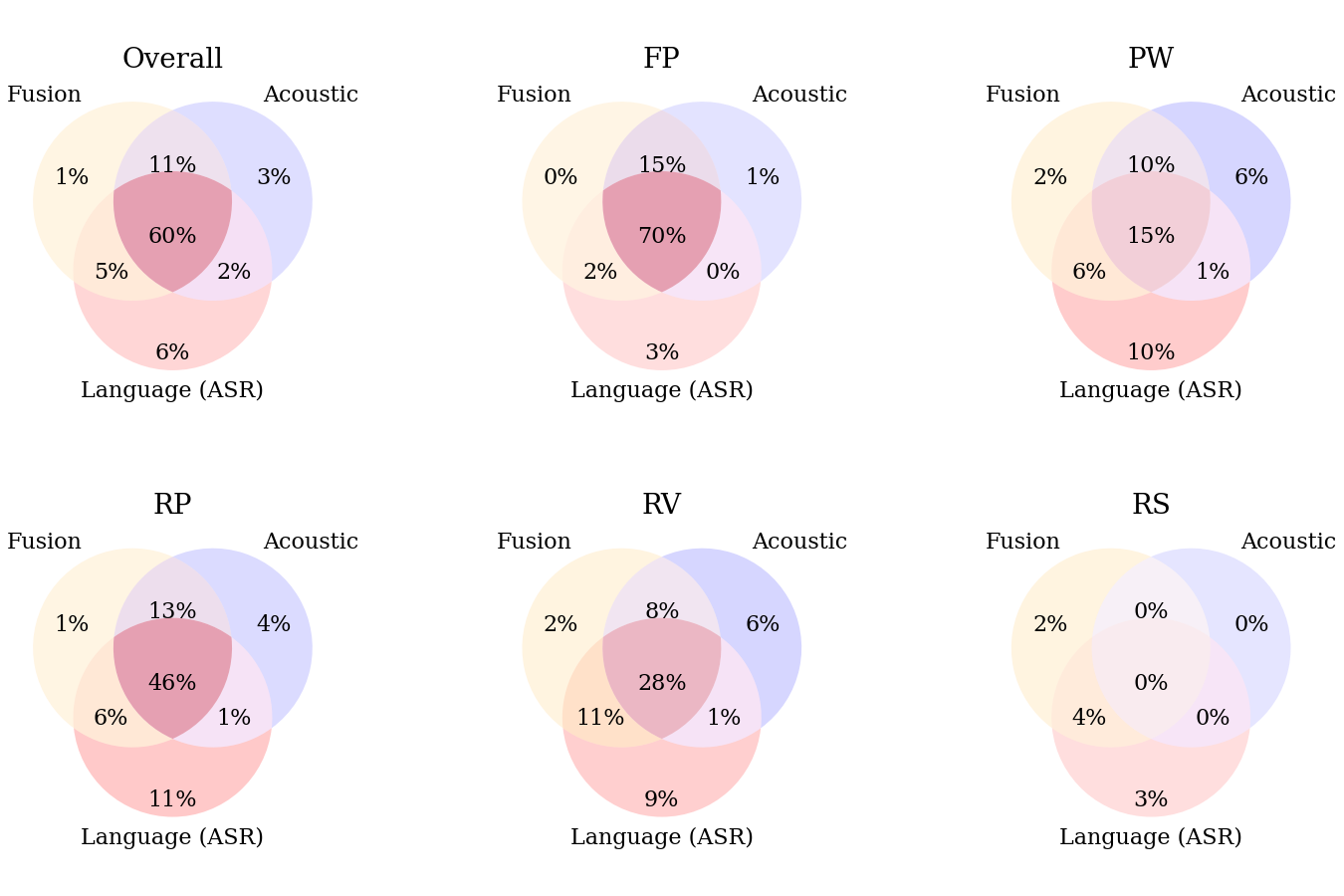}
    \caption{The overlap in frames accurately categorized by the acoustic, language (ASR), and fusion networks. These results are displayed for disfluencies overall, and each disfluency class: filled pauses (FP), partial words (PW), repetitions (RP), revisions (RV), and restarts (RS). For example, for the overall disfluencies diagram, the 60\% in the center indicates that 60\% of all disfluent frames were accurately categorized by all of the three approaches. The color saturation of each patch corresponds to values in each patch.}
    \label{fig:venndiagram}
\end{figure*}

\subsection{Multimodal approaches}

Tables \ref{tab:mm_f1score} and \ref{tab:mm_recallscore} show the multimodal fusion results, where the multimodal models tend to outperform the unimodal approaches. The single-layer perceptron outperforms the unimodal approaches in terms of unweighted and weighted F1 scores, but lags slightly in terms of recall. The BLSTM outperforms the unimodal approaches in terms of all of these metrics and the transformer falls slightly short of the BLSTM. We expect the BLSTM outperforms the transformer for this task because alignment corrections rely most heavily on nearby frames and this is most easily learned by the BLSTM. 

Comparing the multimodal BLSTM to the unimodal approaches, we find improvements in recall and F1 score for most disfluency classes. The largest improvements are for the F1 score of repetitions, revisions, and partial words. For repetitions, the language and acoustic models achieved F1 scores of 0.48 and 0.64, respectively, while the BLSTM achieved an F1 score of 0.69. For revisions, the language and acoustic models achieved F1 scores of 0.45 and 0.44, respectively, while the BLSTM achieved an F1 score of 0.55. For partial words, the language and acoustic models achieved F1 scores of 0.17 and 0.32, respectively, while the BLSTM achieved an F1 score of 0.40. These results suggest that the BLSTM architecture can best leverage the language and acoustic representations.  

Figure \ref{fig:venndiagram} illustrates the overlaps between the acoustic, language, and BLSTM fusion approaches in terms of which disfluent frames they correctly categorized. The values in the yellow circles, for example, show which percentages of frames were correctly categorized by the BLSTM fusion model. Looking at the ``Overall'' subfigure (top-left), 11\% of disfluent frames were only correctly categorized by the fusion and acoustic models, 5\% were only correctly categorized by the fusion and language models, 60\% were correctly categorized by all three, and 1\% were only correctly categorized by the fusion model. These values sum to 77\%, indicating that the fusion model correctly categorized 77\% of disfluent frames. In general, the values in the yellow circles sum to more than the values in the blue (acoustic) circles or red (language) circles individually, indicating that the BLSTM fusion approach outperforms the unimodal approaches. Additionally, looking at the subfigures for partial words, revisions, and restarts, the fusion model accurately categorizes 2\% of frames for each class that were not accurately categorized by the acoustic or language approaches.

\section{Discussion}

Our findings demonstrate that with manually transcribed text, a BERT model can detect and categorize disfluencies with high accuracy. With scalability in mind and ASR transcripts during test time, disfluency detection performance is largely limited by the quality of transcripts and alignments. In terms of the off-the-shelf models, Azure performed the best in transcribing nondisfluent and disfluent speech. After fine-tuning though, Whisper-FT performed better than Azure in transcribing disfluent speech. However, comparing manually transcribed text versus Whisper-FT text as input to the BERT disfluency detection model, we found that performance drops considerably. The most notable drops are for the detection of filled pauses, repetitions, and partial words. 

Our work highlights the potential of acoustic-based disfluency detection. These methods avoid the transcription bottleneck and achieve high accuracy for most disfluency classes. Specifically, we found that WavLM outperformed W2V2 and HuBERT. And in comparing WavLM to the Whisper-FT language approach, we found that WavLM achieves higher F1 scores for the detection of filled pauses, repetitions, revisions, and partial words. 

Finally, our findings show that a multimodal model can offer an improvement over the unimodal approaches. Of the architectures explored, a BLSTM fusion model performed best. Compared to the unimodal models, the BLSTM fusion model yielded the largest improvements for F1 score of repetitions, revisions, and partial words. However, the limitation of the multimodal model is that it does require ASR transcription as well as the loading of both unimodal models. In resource constrained environments, the acoustic approach may be sufficient, but the BLSTM fusion approach offers the best results. 

\section{Conclusion}
\label{sec:conclusion}

In this paper, we provide the first comparison of how different ASR systems and acoustic models could be leveraged for frame-level disfluency detection and categorization. We find that acoustic models outperform the ASR language approaches, but that for most classes, a BLSTM fusion network offers the best performance. In the long term, these findings will lead to more scalable and accurate disfluency detection and categorization models. 

\section{Acknowledgements}
We thank Keli Licata and Elise Jones for providing valuable insight into speech language pathology neeeds for disfluency detection, categorization, and localization. This research is based in part upon work supported by the National Science Foundation (NSF IIS-RI 2006618). This research is supported in part through computational resources and services provided by Advanced Research Computing (ARC), a division of Information and Technology Services (ITS) at the University of Michigan, Ann Arbor.

\vfill\pagebreak

\label{sec:refs}
\bibliographystyle{plain}
\bibliography{main}
\newpage 

\begin{IEEEbiography}
[{\includegraphics[width=1in,height=1in,clip,keepaspectratio]{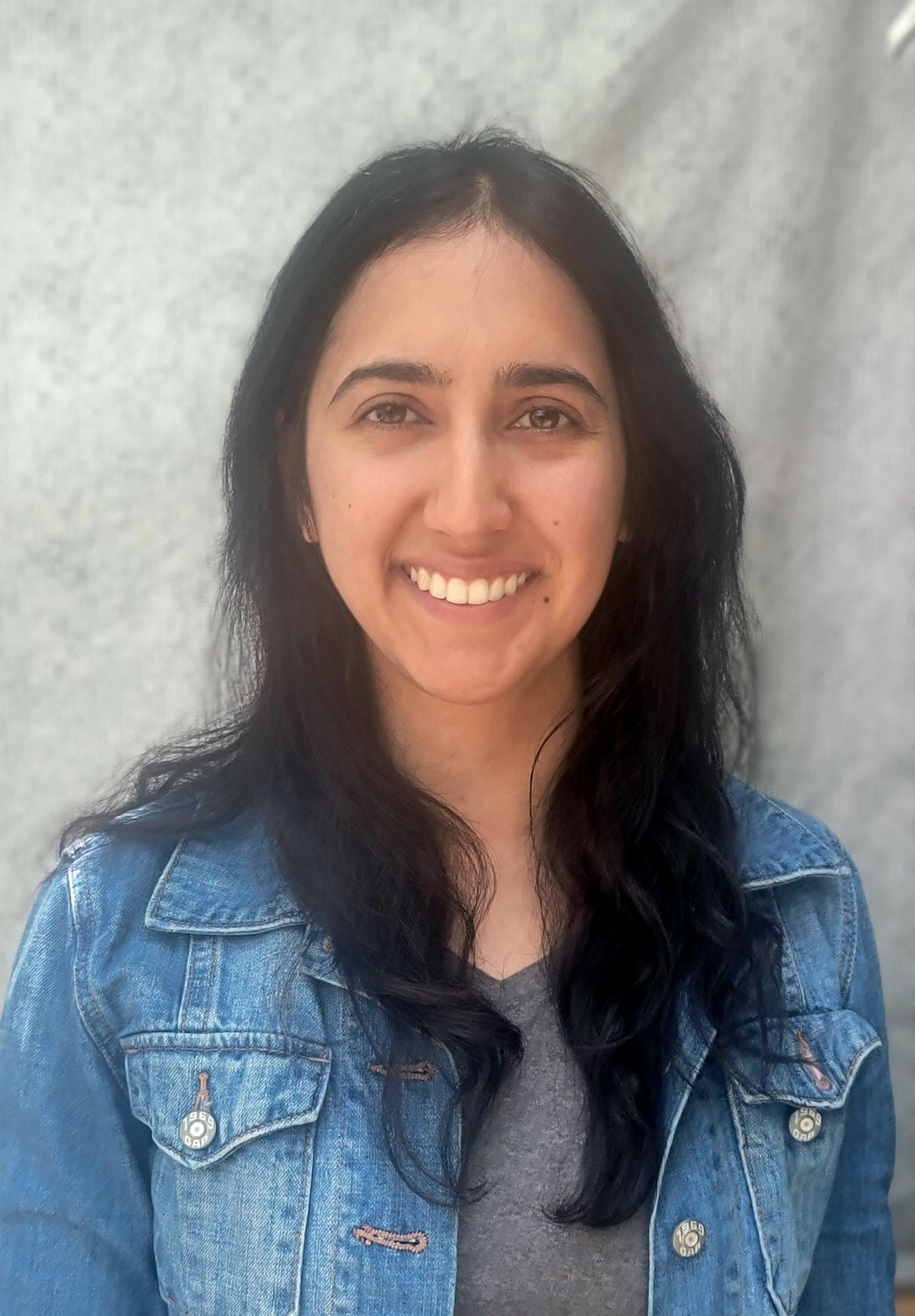}}]{Amrit Romana} received her B.S. degree in Mathematics, and M.S. degree in Computer Science and Engineering from University of Michigan, in 2014 and 2020, respectively. She is currently a Ph.D. student in Computer Science and Engineering at University of Michigan. Her research interests include speech processing and machine learning with the goal of making speech-based technologies more accessible. She is a member of IEEE and ISCA.
\end{IEEEbiography}

\begin{IEEEbiography}
[{\includegraphics[width=1in,height=1in,clip,keepaspectratio]{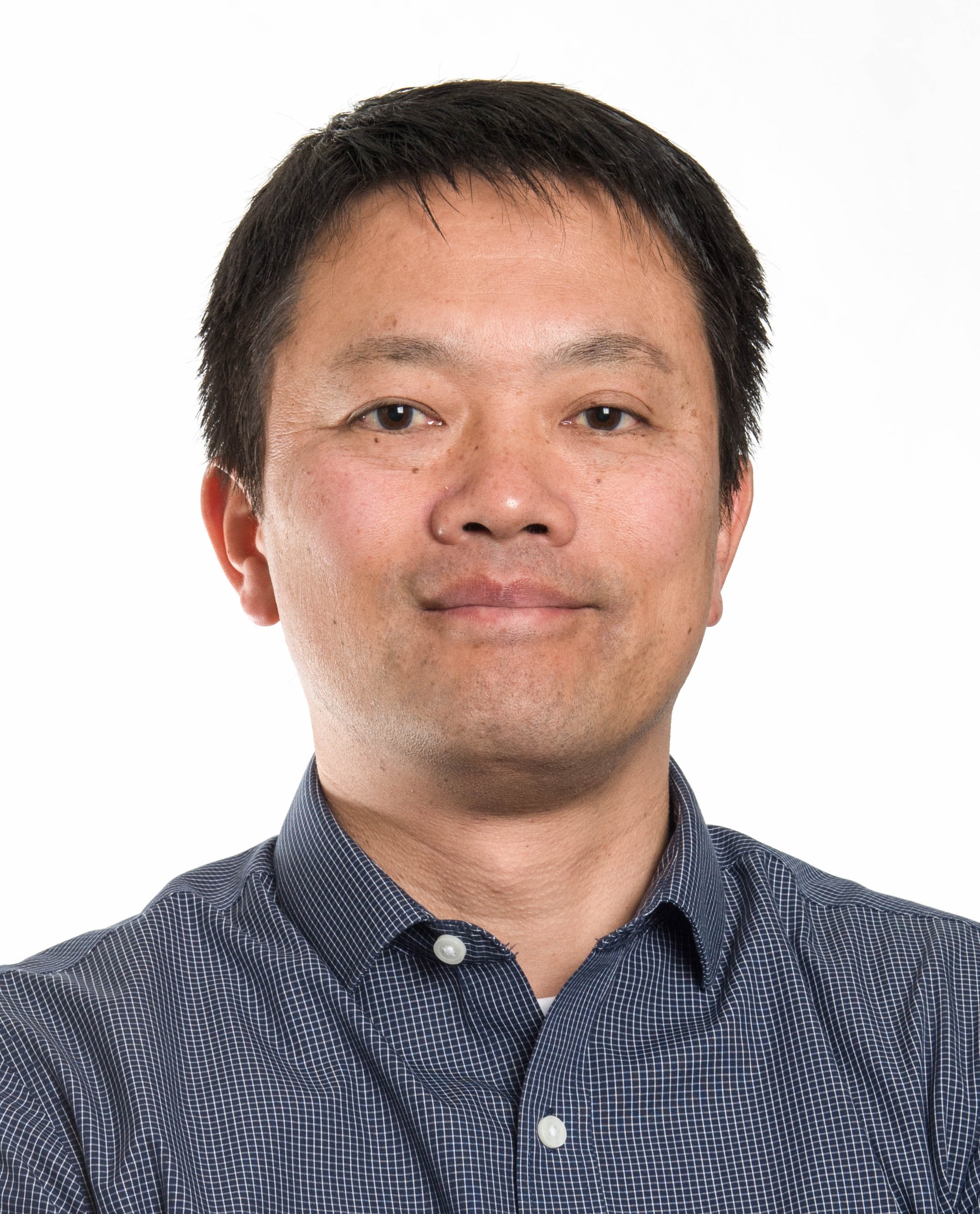}}]{Kazuhito Koishida} received his B.E. degree in electrical and electronic engineering, and M.E., and Dr.Eng. degrees in intelligence science from Tokyo Institute of Technology, Tokyo, Japan, in 1994, 1995 and 1998, respectively. From 1998 to 2000 he was a Post-doctoral Researcher with Signal Compression Laboratory, University of California, Santa Barbara. He joined Microsoft Corporation, Redmond, USA, in 2000 and is currently a Principal Research Manager in Applied Sciences Group. His research interests include speech and audio processing, multimodal signal processing, and machine learning. He is a member of IEEE and ISCA. 
\end{IEEEbiography}

\begin{IEEEbiography}
[{\includegraphics[width=1in,height=1in,clip,keepaspectratio]{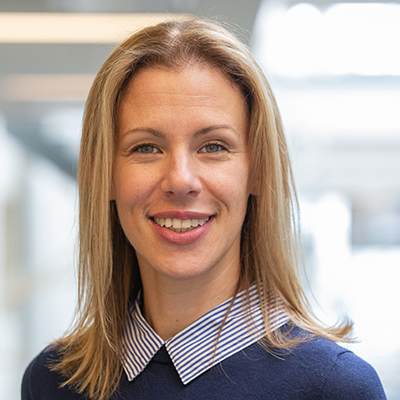}}]{Emily Mower Provost} (M’11, SM'17) is a Professor in Computer Science and Engineering at the University of Michigan. She received her Ph.D. in Electrical Engineering from the University of Southern California (USC), Los Angeles, CA in 2010. She is a Toyota Faculty Scholar (2020) and has been awarded a National Science Foundation CAREER Award (2017), the Oscar Stern Award for Depression Research (2015), a National Science Foundation Graduate Research Fellowship (2004-2007).  She is an Associate Editor for IEEE Transactions on Affective Computing and the IEEE Open Journal of Signal Processing.  She has also served as Associate Editor for Computer Speech and Language and ACM Transactions on Multimedia.  She has received best paper awards or finalist nominations for Interspeech 2008, ACM Multimedia 2014, ICMI 2016, and IEEE Transactions on Affective Computing.  Among other organizational duties, she has been Program Chair for ACII (2017, 2021), ICMI (2016, 2018).  Her research interests are in human-centered speech and video processing, multimodal interfaces design, and speech-based assistive technology. The goals of her research are motivated by the complexities of the perception and expression of human behavior. 
\end{IEEEbiography}

\end{document}